\documentclass[epjST]{svjour3}
\usepackage{cite}
\usepackage{graphics}
\usepackage{graphics}
\begin{document}
\title{Gravity Resonance Spectroscopy \\ and Dark Energy Symmetron Fields}
\subtitle{\textit{q}\textsc{BOUNCE} experiments performed with Rabi and Ramsey Spectroscopy\\ }
\author{Tobias Jenke\inst{1} \and Joachim Bosina\inst{2} \and Jakob Micko\inst{1,2} \and Mario Pitschmann\inst{2}\and Ren$\acute{e}$ Sedmik\inst{2} \and Hartmut Abele\inst{2}}
\institute{Institut Laue-Langevin, 71 avenue des Martyrs, 38000 Grenoble,  France \and Atominstitut TU Wien, Stadionallee 2, 1020 Wien, Austria}
\abstract{Spectroscopic methods allow to measure energy differences with unrivaled precision. In the case of
gravity resonance spectroscopy, energy differences of different gravitational states are measured
without recourse to the electromagnetic interaction. This provides a very
pure and background-free look at gravitation and topics related to the central problem of dark energy and dark matter  at short distances.
In this article we analyse the effect of dark energy scalar symmetron fields, a leading candidate for a screened dark energy field,
and place limits in a large volume of parameter space.} 

\maketitle
\section{Introduction}
\label{intro}
\begin{figure}
	\resizebox{0.5\columnwidth}{!}{\includegraphics{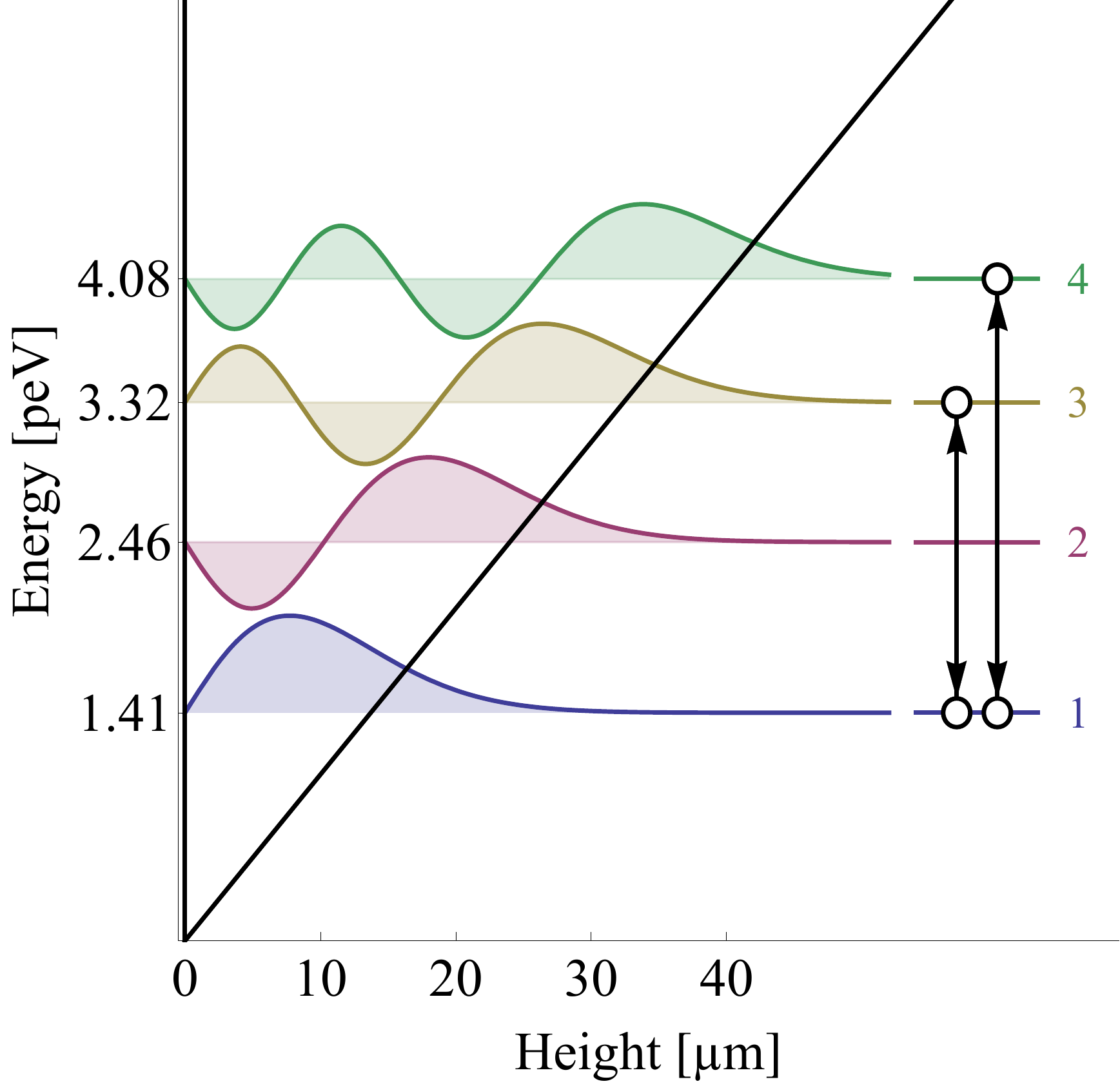}}
	\caption{Energy eigenvalues and eigenfunctions of a neutron above a reflecting mirror bound in the gravity potential of the Earth. By oscillating the mirror with a frequency, which corresponds to the energy difference between quantum states, transitions are induced. The oscillation frequencies are in the acoustic frequency range.}
	\label{fig:1}       
\end{figure}
This article focuses on acoustic Rabi- and Ramsey-transitions between gravitational energy eigenstates of an ultra-cold neutron trapped above a flat neutron reflector in the gravity potential of the Earth.   The method in use is Gravity Resonance Spectroscopy (GRS)~\cite{Jenke:2011}, a method~\cite{Abele:2010,Jenke:2011,Jenke:2014,Cronenberg:2018a} developed by the \emph{q}\textsc{Bounce} collaboration. The energy difference between quantum states in the gravity potential can be related to the frequency of a mechanical modulator, in analogy to the Nuclear Magnetic Resonance technique, where the Zeeman energy splitting of a magnetic moment in an outer magnetic field is connected to the frequency of a radio-frequency field. The frequency range in GRS used so far is in the acoustic frequency range between 100 Hz and 1000 Hz. The quantum states have peV energy, on a much lower energy scale compared to other bound quantum systems.

We present the \emph{q}\textsc{Bounce} experiment and the application of Rabi and Ramsey spectroscopy to GRS in section~\ref{sec:2}.  The lowest stationary quantum state of neutrons in the gravitational field was identified~\cite{Nesvizhevsky:2002a,Nesvizhevsky:2005,Westphal:2007} in 2002 and triggered new experiments and activity in this direction, see e.g.~\cite{Abele:2008,Kreuz:2009,Jenke:2009}.

\emph{q}\textsc{Bounce} started to use  GRS in a search for hypothetical gravity-like interactions. Any deviation from the gravity potential of the Earth would provoke an energy shift in resonance spectroscopy. The \textit{q}\textsc{Bounce} collaboration pioneered searches for dark energy chameleon fields by GRS~\cite{Jenke:2014}. Other limits on chameleon fields stem from neutron interferometer measurements~\cite{Lemmel:2015,TheINDEXCollaboration:2016} and atom interferometry~\cite{Hamilton:2015}, see also~\cite{Jaffe:2017,Sabulsky:2019}, and a living review~\cite{Burrage:2018}. A dark matter axion-like spin-mass coupling limit can be found here ~\cite{Jenke:2014}.  A search for a hypothetical electric charge of the neutron~\cite{Durstberger-Rennhofer:2011} can be provided by GRS as well.

In this presentation we further analyse the effect of additional dark energy scalar symmetron fields~\cite{Hinterbichler:2010}, a leading candidate for a screened dark energy field, which has escaped experimental detection
so far~\cite{Jaffe:2017,Cronenberg:2018a,Sabulsky:2019} and \cite{Burrage:2018}. Details can be found in section~\ref{sec:3}.  For further information on exact symmetron solutions and neutron screening see ref.~\cite{Brax:2018}.


\section{Gravity Resonance Spectroscopy Technique}
\label{sec:2}
\begin{figure}
	\resizebox{0.8\columnwidth}{!}{ \includegraphics{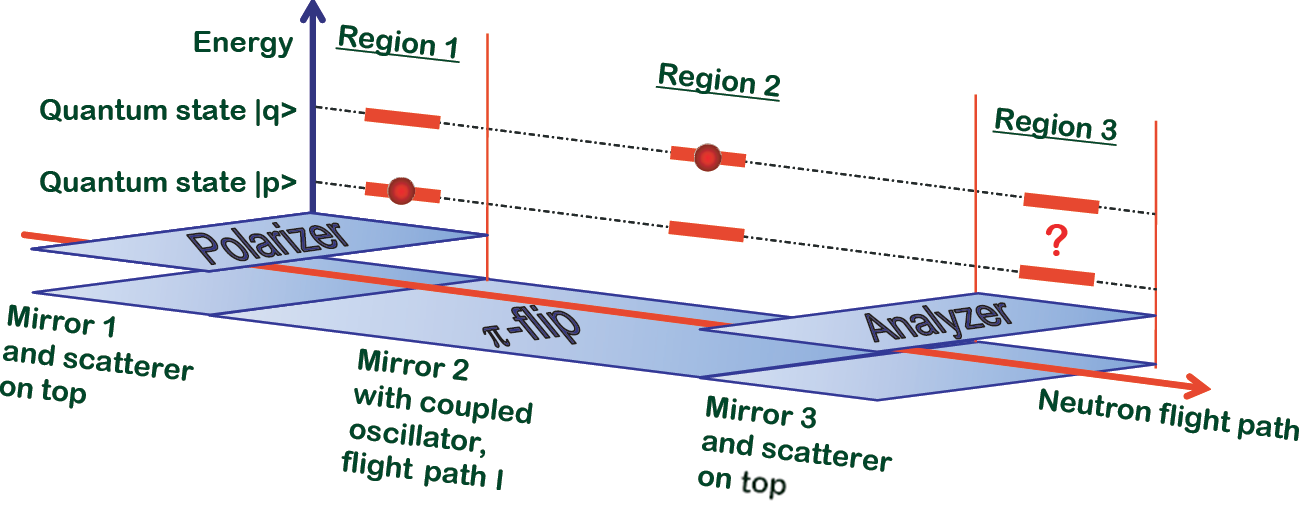}}
	\resizebox{1.\columnwidth}{!}{\includegraphics{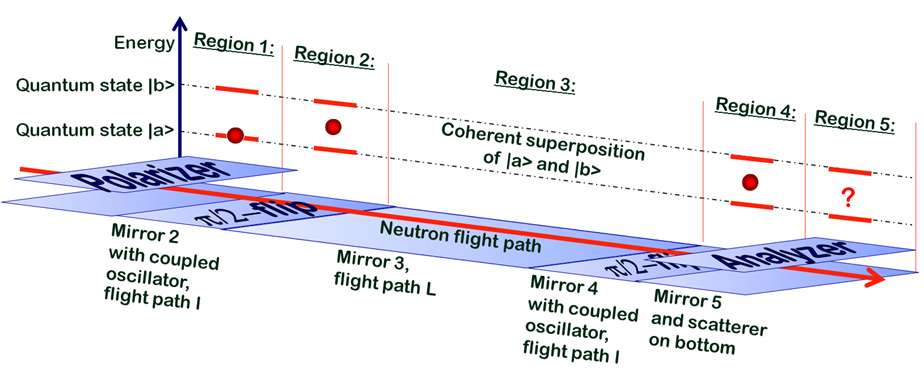}}
	\caption{Schematic views of the experimental set-ups for Rabi (top) and Ramsey (bottom) spectroscopy. Ultra-cold neutrons (UCNs) pass the set-up from left to right.
		Top: Rabi setup. In region I, they are prepared in the gravitational ground state by passing a gap between a rough surface on top and a perfectly polished surface on bottom. Higher states interact with the rough surface, and are effectively scattered off the system. In region II, transitions between quantum states are induced. The surface for that purpose oscillates with variable frequency and strength. This oscillating boundary condition triggers transitions to higher states, if the resonance condition is met, and the oscillation strength is sufficient. Region III is identical to region I and serves again as a state selector.
		Bottom: Ramsey setup: The setup is extended by two more regions. After the state selection in region I, region II puts a neutron into a superposition of two gravitational energy eigenstates. Region III is a time-of-flight region. Here phase manipulations can be done. 
Region IV oscillates in phase with region II according to the principle of Ramsey spectroscopy. Region V is a second state selector.
	}
	\label{fig:2}       
\end{figure}

The \emph{q}\textsc{Bounce} experiment makes use of ultra-cold neutrons with the highest flux worldwide at instrument PF2 at the Institute Laue-Langevin (ILL). The  experimental setup is optimized to the peak velocity of the corresponding PF2 beam line, which is roughly 8m/s. A summary of the \emph{q}\textsc{Bounce} activities can be found in~\cite{Jenke:2019}.
The \emph{q}\textsc{Bounce} collaboration has applied so far the Rabi and the Ramsey method to GRS, see Figure~\ref{fig:2}.  Ultra-cold neutrons (UCNs) pass the set-up from left to right. A highly efficient neutron detector with low background counts the transmitted neutrons. The vertical quantum energy states of the neutrons are quantized in the potential well created by flat glass mirrors and the linear gravitational potential. State selection is provided by rough upper mirrors~\cite{Westphal:2007}. The boundary conditions are realized by glass mirrors with rough or perfectly polished  surfaces.  All mirrors are mounted on nano-positioning tables. An optical system controls the induced mirror oscillations. A movable system based on capacitive sensors  controls and levels steps between the regions. The experiment is shielded by \textbf{$\mu$}-metal against the magnetic field of the Earth. Flux-gate magnetic field sensors log residual magnetic fields. The whole set-up is placed in vacuum.

Regarding Rabi spectroscopy our setup consists of three regions, see Figure~\ref{fig:2}, top. In region one, we use a neutron mirror on top with a rough surface, which acts as a state selector preferring the ground state $|$1$\rangle$. Neutrons in higher states are effectively scattered off the system~\cite{Westphal:2007}. In the second region, we apply sinusoidal mechanical oscillations with tuneable acoustic frequency and amplitude to the neutron mirror to examine the transitions $|p\rangle \rightarrow |q\rangle$. In \cite{Cronenberg:2018a} the transitions $|1\rangle \rightarrow |3\rangle$ and $|1\rangle \rightarrow |4\rangle$ as shown in Figure~\ref{fig:1} are studied. While the original Rabi experiment and all subsequent resonance experiments have been using electromagnetic interactions to drive the excitations, here, mechanical oscillations of the neutron mirror are used. Alternatively, magnetic gradient fields have been proposed~\cite{Kreuz:2009}. The final region is realized by a second state selector followed by a neutron-counting detector.

Regarding Ramsey spectroscopy~\cite{Sedmik:2019}, we have added two more regions to allow a long time of flight region III for a neutron being in a superposition between state $q$ and state $p$, after and before passing  a so-called $\pi$/2 region, see Figure~\ref{fig:2}, top. Ramsey spectroscopy has several advantages over Rabi spectroscopy, like improvements in statisticial accuracy.

\section{Dark Energy Symmetron Fields and Restrictions from GRS}
\label{sec:3}

General relativity delivers a standard framework for describing the structure of the cosmos, from the Big Bang to the current universe undergoing an accelerated expansion. The source $T_{\mu \nu}$ must involve a certain number of postulated ingredients: an inflaton
field, the matter of the Standard Model, a dark matter component, and a cosmological constant contribution often described as vacuum energy. The small magnitude of the vacuum energy is troubling from a theoretical point of view and an appealing resolution to this problem is the introduction of additional scalar fields. However, these have so far escaped experimental detection, suggesting some kind of screening mechanism may be at play. A natural screening mechanism based on spontaneous symmetry breaking has been proposed, in the form of symmetrons~\cite{Hinterbichler:2010}. Symmetron dark energy is a simple field theoretic model, which captures most of the features of screened modified gravity. The potential of the real scalar field $\phi$, which was named symmetron because of its Z2 symmetry $\varphi$ $\rightarrow$ -$\varphi$, reads

\begin{equation}
\mathcal{V}(\varphi) = -\frac{\mu^2}{2} \varphi^2 + \frac{\lambda}{4} \varphi^4 + \frac{\mu^4}{4 \lambda},
\end{equation}
where $\sqrt{-\mu^2)}$
is the mass analogous as for the Higgs mechanism and $\lambda$ a dimensionless positive constant describing self-interaction. In vacuum, the minimum energy configuration breaks the Z2 symmetry and the field takes on a vacuum expectation value VEV($\varphi$)= $\pm \varphi_V=\mu/\sqrt{\lambda}$. In the equation above the coupling to matter, which must also respect the Z2 symmetry, is not yet included. Incorporating the symmetron-matter coupling leads to the effective potential in a medium with mass density $\rho$= $\rho/(\hbar^3c^5)$, which is given by
\begin{equation}
\mathcal{V}_{eff}(\varphi) \sim V(\varphi)  +\frac{\rho}{2M^2}\varphi^2+\frac{\mu^4}{16 \pi^2}~\frac{\lambda}{4}\varphi^4+(\frac{\rho}{2M^2}-\frac{\mu^2}{2})\varphi^2+\frac{\mu^4}{4\lambda}+\frac{\mu^4}{16 \pi^2},
\end{equation}
where $M$ is an inverse coupling parameter to matter of dimension energy. The quantum fluctuations of the symmetron field lift the minimum of the potential by an amount proportional to $\delta V \sim \mu^4/(16\pi^2)$ playing the role of dark energy and leading to the cosmic acceleration. However, these fluctuations neither affect the symmetron fields nor the experimental observables. The field acquires a nonzero VEV in low-density regions, whilst the symmetry is restored in high-density regions. In the latter, the field effectively disappears and is consequently unobservable. In regions of low density, the field spontaneously breaks symmetry and acquires a non-vanishing VEV. In this case, it couples to matter and mediates a fifth force, as is the case in our experimental vacuum.
Such fields would change the energy of quantum states of ultra-cold neutrons. Our spectroscopic approach based on the Rabi resonance method for section~\ref{sec:2} probes these quantum states with a resolution of $\Delta$ E = 2$\times$$10^{-15}$eV. This allows us to exclude the symmetron as the origin of dark energy for a large volume of the three-dimensional parameter space.

\begin{figure}[h]
	\resizebox{0.8\textwidth}{!}{\includegraphics{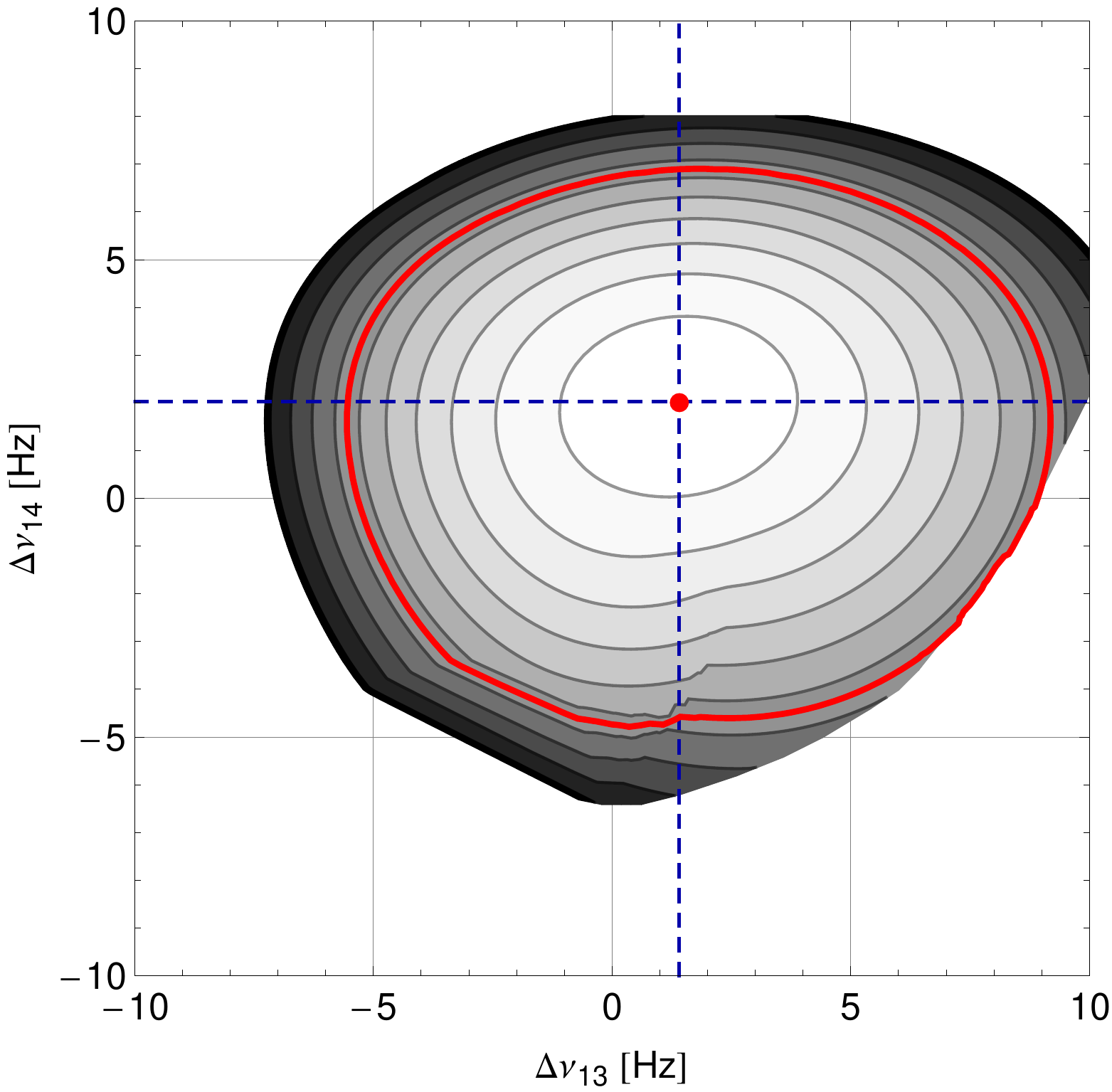}}
	\caption{$\chi^2$-hypercontour with respect to $\Delta\nu_{13}$ and $\Delta\nu_{14}$: The global minimum $\chi^2_{min}$ is found at $\Delta\nu_{13} = 1.402$~Hz and $\Delta\nu_{14} = 2.026$~Hz. The contour lines are equally spaced with a distance of $\Delta\chi^2 = 1$. The red curve shows the contour $\Delta\chi^2_{up} = 6.25$ which corresponds to a confidence level of 90\% for the three additional (positive!) symmetron parameters.}
	\label{fig:3}       
\end{figure}
In the following, the extraction of experimental bounds is summarized:
In~\cite{Cronenberg:2018a}, we used the predicted shifts of the resonance frequencies for the measured transitions $|1\rangle \leftrightarrow |3\rangle$ and $|1\rangle \leftrightarrow |4\rangle$, added the symmetron model parameters $\mu$, $\lambda$ and $M$ as fit parameters, fixed the Earth’s acceleration $g$ to the local value of $g_{Grenoble}$, and performed a full $\chi^2$-analysis. As a result, the extracted limits take into account all possible correlations between the fit parameters, including the symmetron model parameters. The disadvantage of this method is its complexity. Therefore, the extension of the excluded parameter space is done slightly differently: In a first step, the experimental data is fitted using the transition frequencies $\nu_{13}^{fit} = \nu_{13}(g = g_{Grenoble}) + \Delta\nu_{13}$ and $\nu_{14}^{fit} = \nu_{14}(g = g_{Grenoble}) + \Delta\nu_{14}$. The corresponding matrix overlap integrals are parametrized using $\nu_{13}^{fit}$ and $\nu_{14}^{fit}$. The two free fit parameters $\Delta\nu_{13}$ and $\Delta\nu_{14}$ correspond to the deviations from the predicted state transition frequencies in Grenoble. The 2D-$\chi^2$-hypercontour with respect to $\Delta\nu_{13}$ and $\Delta\nu_{14}$ is extracted. To do so, the fit is repeated on a grid for fixed values of $\Delta\nu_{13}$ and $\Delta\nu_{14}$, while all other fit parameters (the count rate, and the contrasts) are varied. Then, contours of these hypercontours are calculated. The result is shown in Fig.~\ref{fig:3}.

\begin{figure}
	\resizebox{\textwidth}{!}{\includegraphics{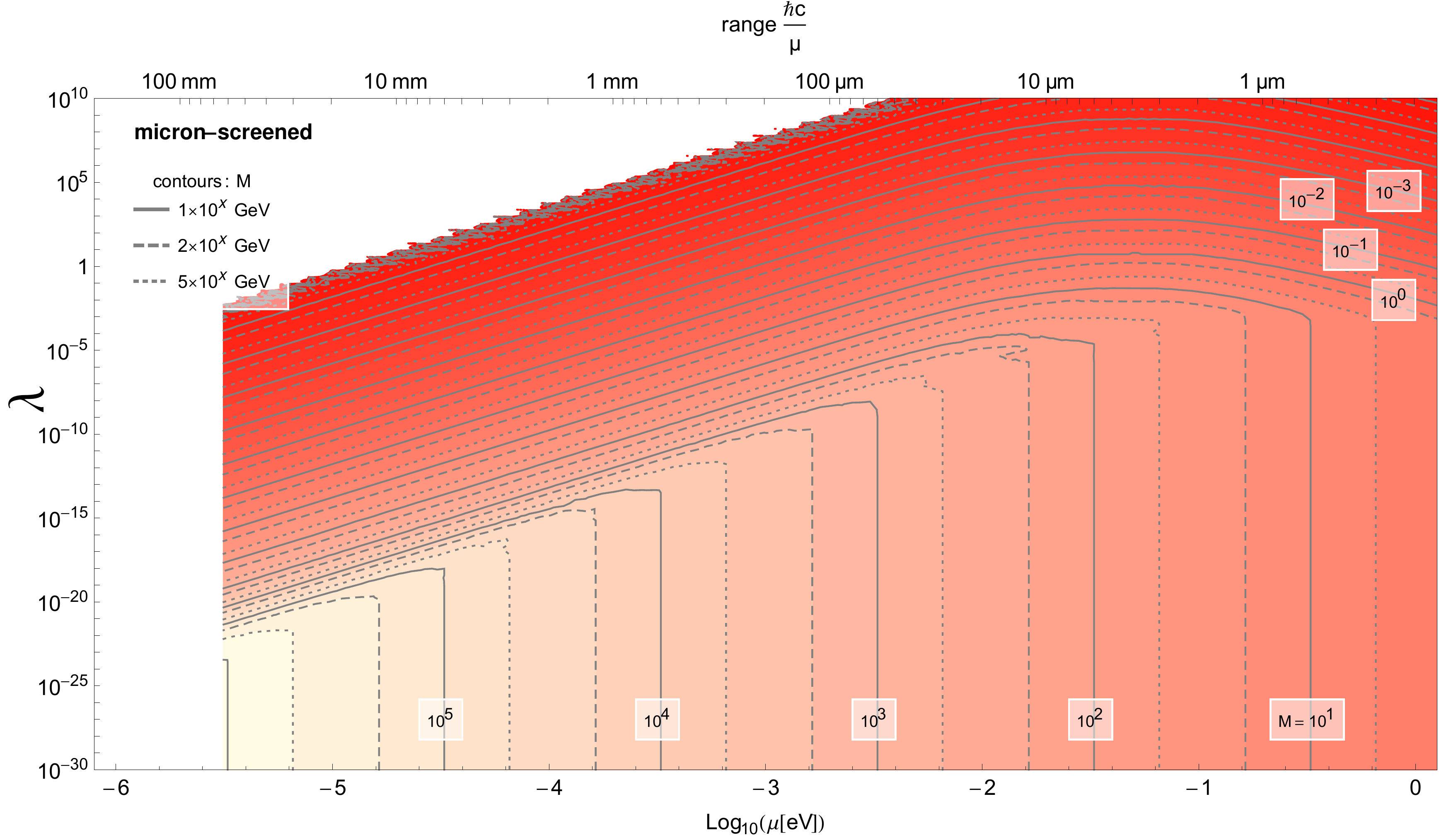}}
	\resizebox{\textwidth}{!}{\includegraphics{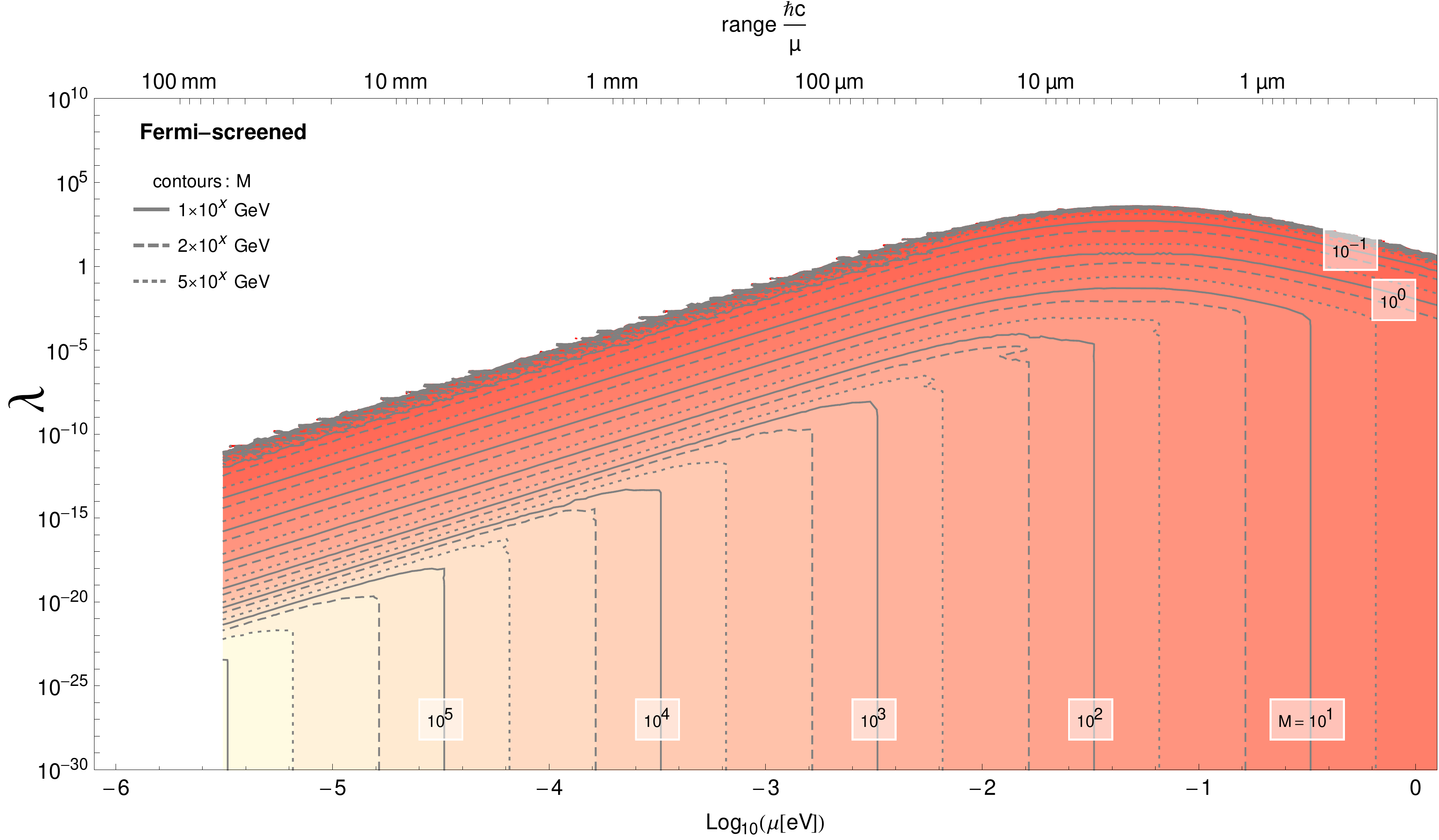}}
	\caption{Experimental limits on the existence of dark energy symmetron fields: The contours show the full information of the excluded 3D parameter space: For given values of the bare potential mass scale $\mu$ and the self-coupling $\lambda$, all values of the inverse coupling strength~$M$ smaller than indicated in the contour are excluded with a confidence of $90\%$. This confidence level corresponds to the probability contents inside a hypercontour of $\chi^2 =
\chi^2_{min} + 6.25$. In both figures, the screening factor~$Q$ is taken into account, in which the radius $R_N$ corresponds to the neutron size as a source for the symmetron (for additional explications, see \cite{Brax:2018} and supplementary information in \cite{Cronenberg:2018a}). In the upper figure ("micron-screening"), $R_N$ is taken as $z_0=5.9~{\mu}$m, which corresponds to the vertical extention of its Schrödinger wave function. In the lower figure ("Fermi-screening"), $R_N$ is taken as 0.5~fm.}
	\label{fig:4}       
\end{figure}
\begin{figure}
	\resizebox{0.8\textwidth}{!}{\includegraphics{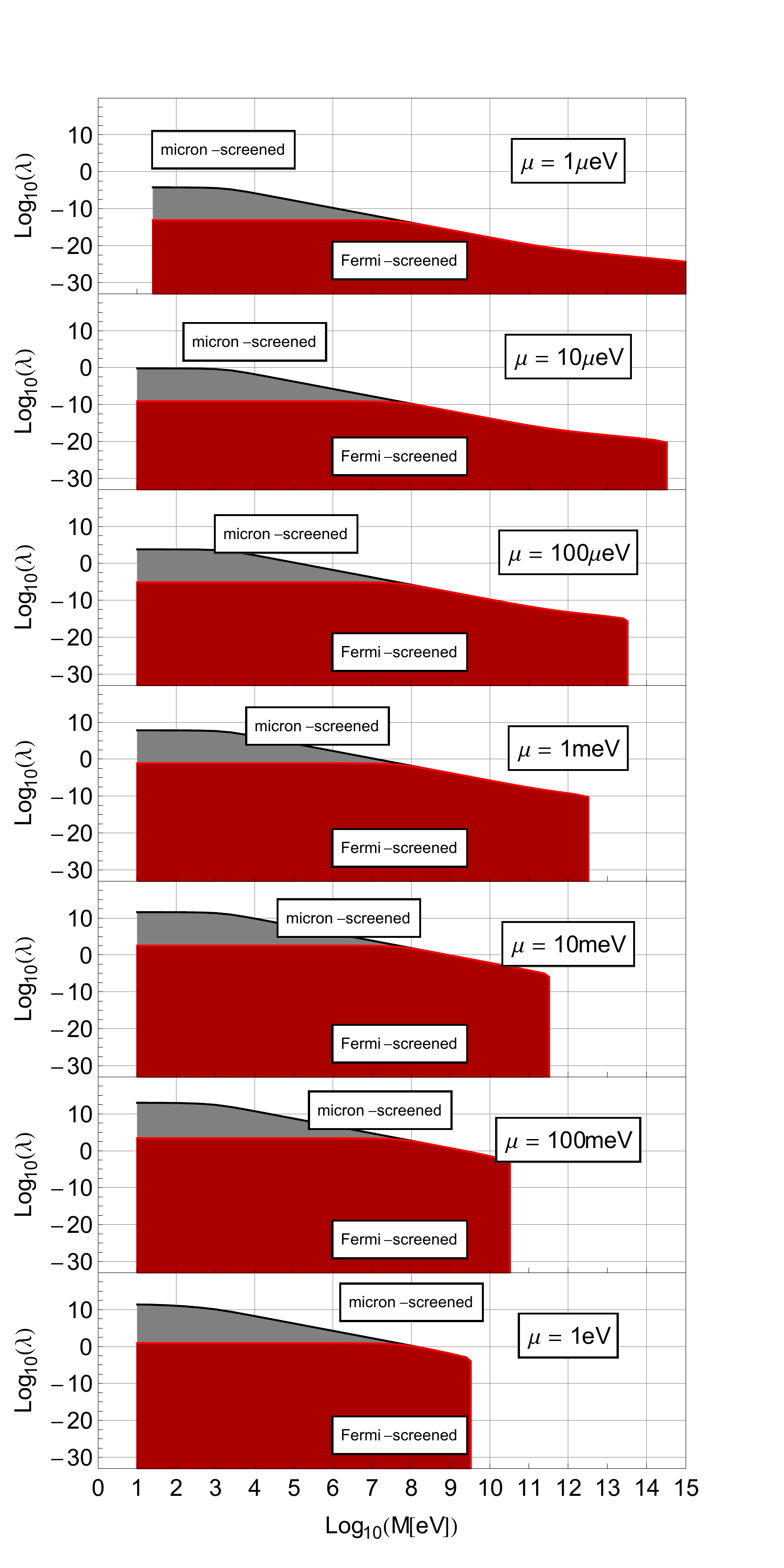}}
	\caption{This figure presents vertical cuts in the excluded 3D parameter space of Fig.~\ref{fig:3}. Here, the two limiting cases \textit{micron}- and \textit{fermi}-screened are shown in gray and red.}
	\label{fig:5}       
\end{figure}
As a second step, the predicted frequency shifts due to the symmetron field are calculated taking into account the screening function and boundary conditions due to finite size of the vacuum chamber.  We treat the symmetron as a classical field, and employ a semi-classical description for the neutron acting as source of the symmetron. A detailed calculation of the symmetron prediction is found in the Supplementary Information in \cite{Cronenberg:2018a} and \cite{Brax:2018}.
To derive experimental limits at a confidence level of 90\%, the points of intersections $\Delta\nu_{13}^{Symmetron}(\mu,\lambda,M) = \Delta\nu_{13}(\Delta\chi^2 = \Delta\chi^2_{up})$ and $\Delta\nu_{14}^{Symmetron}(\mu,\lambda,M) = \Delta\nu_{14}(\Delta\chi^2 =\Delta\chi^2_{up})$ are taken. We set $\Delta\chi^2_{up} = 6.25$, which corresponds to the number of free fit parameters taking into account the positive sign of the symmetron model parameters. The confidence level resembles to the probability contents inside a hypercontour of $\chi^2 =
\chi^2_{min} + 6.25$.
The full result is shown in Fig.~\ref{fig:4}: Regarding the screening, we consider two limiting cases. First, in the upper figure, we assume that the Schrödinger probability density is the source of the symmetron as motivated by the Schrödinger-Newton equation~\cite{Moroz:1998,Penrose:1998,Bahrami:2014,Brax:2018} ({\it{micron}}-screening). For the other limiting case ({\it{fermi}}-screening), shown in the lower figure, we assume that the well-defined size of the neutron provided by the quark-gluon dynamics and quantified by its form factor describes the interaction region with the symmetron.

Finally, we present vertical cuts in the 3D-parameter space of Fig.~\ref{fig:4} in Fig.~\ref{fig:5}. Here, the two limiting cases {\it{micron}}- and {\it{fermi}}-screened are shown in different colors.

\end{document}